\documentclass[12pt]{article}
\usepackage{a4wide,epsfig,psfrag,amsmath,amssymb,cite,scalefnt}
\usepackage{color}

\graphicspath{ {figs/} }

\newcommand{\gsim}{\;\rlap{\lower 3.5 pt \hbox{$\mathchar \sim$}} \raise 1pt
 \hbox {$>$}\;}
\newcommand{\lsim}{\;\rlap{\lower 3.5 pt \hbox{$\mathchar \sim$}} \raise 1pt
 \hbox {$<$}\;}

\begin{document}

\title{\vskip-3cm{\baselineskip14pt
    \begin{flushleft}
      \normalsize TTP15-029
  \end{flushleft}}
  \vskip1.5cm
  $\Gamma(H\to b\bar{b})$ to order $\alpha\alpha_s$
}

\author{
  Luminita Mihaila$^{(a)}$, Barbara Schmidt$^{(b)}$ and Matthias Steinhauser$^{(b)}$
  \\[1em]
  {\small\it (a) Institut f{\"u}r Theoretische Physik, Universit{\"a}t
    Heidelberg, 69120 Heidelberg, Germany }
  \\[1em]
  {\small\it (b) Institut f{\"u}r Theoretische Teilchenphysik,
    Karlsruhe Institute of Technology (KIT)}\\
  {\small\it 76128 Karlsruhe, Germany}  
}
  
\date{}

\maketitle

\thispagestyle{empty}

\begin{abstract}

  We compute the decay rate of the Standard Model Higgs boson to bottom quarks
  to order $\alpha\alpha_s$. We apply the optical theorem and calculate the
  imaginary part of three-loop corrections to the Higgs boson propagator
  using asymptotic expansions in appropriately chosen mass ratios.  The
  corrections of order $\alpha\alpha_s$ are of the same order of magnitude as
  the ${\cal O}(\alpha_s^3)$ QCD corrections but have the opposite sign.

  \medskip


\end{abstract}

\thispagestyle{empty}


\newpage


\section{Introduction}

After the discovery of a Higgs boson in run I of the CERN Large Hadron
Collider it is one of the main tasks of run II to determine the properties of
the new particle. Among them is the coupling to other particles. This is
predominantly done by determining Higgs production cross sections and decay
branching ratios, i.e. the ratio of the partial decay width of the Higgs boson
to the considered particles normalized to the total decay rate.  The latter is
dominated by the partial decay rate to bottom quark, $\Gamma(H\to b\bar{b})$,
which hence influences all branching ratios.  Thus, $\Gamma(H\to b\bar{b})$
should be available as precisely as possible.

QCD corrections are known up to order $\alpha_s^4$ (see, e.g.,
Refs.~\cite{Gorishnii:1990zu,Chetyrkin:1997mb,Chetyrkin:1996sr,Chetyrkin:1997vj,Harlander:1997xa,Baikov:2005rw,Mihaila:2013wma})
and first results of order $\alpha_s^5$ induced by virtual top quarks have
been obtained in Ref.~\cite{Liu:2015fxa}.  Good convergence of the
perturbative series is observed leading to a $0.1\%$ contribution of the
$\alpha_s^4$ corrections to $\Gamma(H\to b\bar{b})$.  As far as electroweak
corrections are concerned only one-loop corrections are available which have
been computed beginning of the
nineties~\cite{Kniehl:1991ze,Dabelstein:1991ky}.  At two- and three-loop
order only the leading $M_t^2$ corrections are
available~\cite{Kwiatkowski:1994cu,Kniehl:1994ju,Chetyrkin:1996wr,Chetyrkin:1996ke}.
In this work we compute QCD corrections to the full ${\cal O}(\alpha)$ result
and thus obtain all contributions of order $\alpha\alpha_s$ to the partial decay
rate of a Standard Model Higgs boson into bottom quarks.
Analog corrections to the decay rates of the $Z$ and $W$ bosons have been
computed in Refs.~\cite{Czarnecki:1996ei,Harlander:1997zb,Fleischer:1999iq}
and~\cite{Kara:2013dua}, respectively.

We parametrize the corrections to the decay rate as follows
\begin{eqnarray}
  \Gamma(H\to b\bar{b}) &=& \Gamma^{(0)}\left(
  1 + \Delta^{(\alpha_s)} + \Delta^{(\alpha)} + \Delta^{(\alpha\alpha_s)} + 
  \ldots
  \right)
  \label{eq::Gam}
  \,,
\end{eqnarray}
where the ellipses stand for higher order corrections in $\alpha$ and
$\alpha_s$.  It is convenient to split the electroweak corrections
into a weak and a QED contributions which to our order are separately
finite and gauge invariant:
\begin{eqnarray}
  \Delta^{(\alpha)} &=& \Delta^{(\rm QED)} + \Delta^{(\rm weak)}
  \,,\nonumber\\
  \Delta^{(\alpha\alpha_s)} &=& \Delta^{(\rm QED,\alpha_s)} 
  + \Delta^{(\rm weak,\alpha_s)}
  \,.
\end{eqnarray}
The aim of this paper is the computation of the mixed
corrections $\Delta^{(\alpha\alpha_s)}$.  In Eq.~(\ref{eq::Gam})
$\Gamma^{(0)}$ denotes the Born decay rate which is given by
\begin{eqnarray}
  \Gamma^{(0)} &=& \frac{N_c \alpha m_b^2 M_H}{8 s_W^2 M_W^2}\beta_0^3
  \,,
  \label{eq::Gam0alpha}
\end{eqnarray}
where $N_c=3$ is the number of colours and 
$s_W$ is the sine of the weak mixing angle.
$\beta_0=\sqrt{1-4m_b^2/M_H^2}$ is the velocity of the produced bottom
quarks which from now on we approximate to $\beta_0=1$.
As an alternative to Eq.~(\ref{eq::Gam0alpha}) one can
replace the fine structure constant by Fermi's constant
via
\begin{eqnarray}
  \frac{G_F}{\sqrt{2}} &=& \frac{\pi\alpha}{2 s_W^2 M_W^2}\frac{1}{1-\Delta r}
  \,,
\end{eqnarray}
where the finite quantity 
$\Delta r$ parametrizes the radiative corrections to the muon
decay beyond QED corrections within the effective four-fermion
theory~\cite{Sirlin:1980nh}. This leads to
\begin{eqnarray}
  \Gamma^{(0)} &=& \frac{N_c G_F m_b^2 M_H}{4\sqrt{2}\pi}
  \,.
  \label{eq::Gam0GF}
\end{eqnarray}

For later reference we provide the Born decay rate including higher order
terms in $\epsilon=(4-d)/2$ which are useful in the renormalization
procedure. For $d\not=4$ both~(\ref{eq::Gam0alpha}) and~(\ref{eq::Gam0GF})
have to be multiplied by the factor\footnote{Throughout this paper we adopt a
  $\overline{\rm MS}$-like convention and set $\gamma_E$ and $\log(4\pi)$ to
  zero.}
\begin{eqnarray}
  f(\epsilon) &=& 
  \left(\frac{\mu^2}{M_H^2}\right)^\epsilon
  \left[1 + \epsilon + \epsilon^2\left(4-{\frac{\pi^2}{4} } \right) 
    + {\cal O}(\epsilon^3) \right]
  \,.
\end{eqnarray}

For later convenience we also list the one-loop QCD
corrections including terms of order~$\epsilon$
\begin{eqnarray}
  \Delta^{(\alpha_s)} &=& \frac{\alpha_s}{\pi} C_F
  \left\{ 
  \frac{17}{4} + \frac{3}{2}\ln\left(\frac{\mu^2}{M_H^2}\right) 
  + \epsilon \left[ 
    \frac{179}{8} - \frac{7\pi^2}{8} - 6\zeta(3) 
    + \frac{23}{2}\ln\left(\frac{\mu^2}{M_H^2}\right) 
    \right.\right.\nonumber\\&&\mbox{}\left.\left.
    + \frac{9}{4}\ln^2\left(\frac{\mu^2}{M_H^2}
    \right)\right]\right\}
  \,,
\end{eqnarray}
where $C_F=4/3$.
To obtain this result the bottom quark mass has been renormalized in the
$\overline{\rm MS}$ scheme.  The one-loop QED corrections are obtained from
$\Delta^{(\alpha_s)}$ with the help of
\begin{eqnarray}
  \Delta^{(\rm QED)} &=&
  \frac{\alpha Q_b^2}{C_F\alpha_s}\Delta^{(\alpha_s)} 
  \,,
\end{eqnarray}
where $C_F=4/3$ and $Q_b=-1/3$ is the charge of the bottom quark.

The remainder of the paper is organized as follows: In the next
Section we discuss the method we want to use for the three-loop
diagrams of order $\alpha\alpha_s$ and apply it to the
one-loop electroweak corrections. The ${\cal O}(\alpha\alpha_s)$
corrections are presented afterwards in Section~\ref{sec::alas}.
In Section~\ref{sec::num} we discuss the numerical effect,
compare with the known QCD corrections and conclude.


\section{\label{sec::al}Corrections of order $\alpha$}

Before discussing the computation of the genuine diagrams of order
$\alpha$ we briefly elaborate on the counterterm contribution.  We
follow Ref.~\cite{Denner:1991kt} and introduce one-loop counterterms for the
Higgs boson wave function ($\delta Z_H$), the vacuum expectation value
($\delta v$) and the bottom quark mass ($\delta_{m_b}$). This leads
to the following counterterm contribution of $\Delta^{(\rm weak)}$ 
\begin{eqnarray}
  \Delta_{\rm CT}^{(\rm weak)} &=&
  \Gamma^{(0)}\left(1 - 2\frac{\delta v}{v} + \delta Z_H - \Delta r +
  2\delta_{m_b}\right)
  \,,
  \label{eq::DeltaalCT}
\end{eqnarray}
where $\Gamma^{(0)}$ is given in Eq.~(\ref{eq::Gam0GF}). 
We do not include the on-shell wave function renormalization of the quarks
in $\Delta_{\rm CT}^{(\rm weak)}$ since it is automatically taken into account
when using the optical theorem (see below). In the on-shell
scheme the mass counterterm is defined through
\begin{eqnarray}
  m_b^0 &=& M_b \left(1 + \delta_{m_b}^{\rm OS}\right)
  \,,
\end{eqnarray}
with $M_b$ being the on-shell mass. We take $\delta_{m_b}^{\rm OS}$
from Ref.~\cite{Hempfling:1994ar,Kniehl:2004hfa} dropping all tadpole
contributions.  The divergent part of $\delta_{m_b}$ determines the
$\overline{\rm MS}$ counterterm $\delta_{m_b}^{\overline{\rm MS}}$ and
is in our approximation (i.e. at most $m_b^2$ terms in the decay rate)
given by
\begin{eqnarray}
  \delta_{m_b}^{(\rm weak), \overline{\rm MS}} &=&
  \delta_b^{\rm OS}\Big|_{1/\epsilon~\mbox{pole}} \,\,=\,\, 
  \frac{\alpha}{\pi}\left(-\frac{3m_t^2}{32s_W^2M_W^2} 
  + \frac{9 +  6s_W^2 -  8s_W^4}{96c_W^2s_W^2}\right)\frac{1}{\epsilon} 
  \label{eq::deltaMBalpha}\,.
\end{eqnarray}
The remaining terms on the right-hand side of
Eq.~(\ref{eq::DeltaalCT}) can be written in the form~\cite{Denner:1991kt}
\begin{eqnarray}
  v_r &\equiv& - 2\frac{\delta v}{v} + \delta Z_H - \Delta r 
  \nonumber\\&=&
  -\frac{\Sigma^W(0)}{M_W^2} - \Sigma^{\prime H}(M_H^2)
  -\frac{2}{s_Wc_W}\frac{\Sigma^{\gamma Z}(0)}{M_Z^2} -
  \frac{\alpha}{4\pi s_W^2}\left(6 +
  \frac{7-4s_W^2}{2s_W^2}\ln(c_W^2)\right)
\nonumber\\
&=&
\frac{\alpha}{ \pi s_W^2 }\bigg\{-\frac{1+2 c_W^2}{8 c_W^2}\frac{1}{\epsilon}
-\frac{3 (1+2 c_W^2)}{32 c_W^2}+\frac{3 (1 +2 c_W^4) M_Z^2}{8 c_W^2 M_H^2}
+\frac{(13-2 \sqrt{3}\pi) M_H^2}{32 M_W^2}
\nonumber
\\
&&+
\frac{(M_H^2 - 6 M_W^2) (M_H^2 + 2 M_W^2)}{4 M_H^3 \sqrt{-M_H^2 + 4 M_W^2}}
\arctan{\frac{M_H}{\sqrt{-M_H^2 + 4 M_W^2}}}
\nonumber
\\
&&+
\frac{(M_H^2 - 6 M_Z^2) (M_H^2 + 2 M_Z^2)}{8 c_W^2 M_H^3 \sqrt{-M_H^2 + 4 M_Z^2}}
\arctan{\frac{M_H}{\sqrt{-M_H^2 + 4 M_Z^2}}}
\nonumber
\\
&&+
\frac{(M_H^2 + 5 M_W^2)}{48 M_W^2}\ln\left(\frac{\mu^2}{M_H^2}\right)
+ \frac{(-1 + 11 c_W^2 + 8 c_W^4)}{48
  c_W^2}\ln\left(\frac{\mu^2}{M_Z^2}\right)
\nonumber
\\
&&+ 
\frac{M_H^2 c_W^2 - M_W^2(5 + 18 c_W^2 + 8 c_W^4)}{48
  M_W^2 c_W^2} 
\ln\left(\frac{\mu^2}{M_W^2}\right)
\nonumber
\\
&&+ 
 \frac{(M_H^4 - 5 M_H^2 M_W^2 - 5 M_W^4)}{48 M_W^2 (M_H^2 - M_W^2)}
\ln\left(\frac{M_W^2}{M_H^2}\right)
\nonumber
+
\frac{(-5 + 8 c_W^2)(1+c_W^2+c_W^4) }{48 c_W^2 s_W^2}\ln\left(\frac{M_W^2}{M_Z^2}\right)
\nonumber
\\
&&+
N_c \bigg[
\frac{m_t^2(8 m_t^2+ M_H^2)}{16 M_H^2 M_W^2 } + 
\frac{m_t^2 (2 m_t^2 + M_H^2)\sqrt{4 m_t^2 - M_H^2}}{4 M_H^3 M_W^2}
\arctan{\frac{M_H}{\sqrt{4 m_t^2 - M_H^2}}}
\bigg]
\bigg\}
\,,
\nonumber\\
  \label{eq::deltaVRa}
\end{eqnarray}
where $v$ is the vacuum expectation value and
$\Sigma^W$, $\Sigma^{\gamma Z}$ and $\Sigma^H$ denote the two-point
functions of the corresponding bosons in the notation given in
Ref.~\cite{Denner:1991kt}. The prime in the case of the Higgs boson two-point
function denotes the derivative w.r.t. the external momentum squared,
$q^2$. Afterwards $q^2=M_H^2$ is chosen.

\begin{figure}[t]
  \begin{center}
    \includegraphics[width=.22\textwidth]{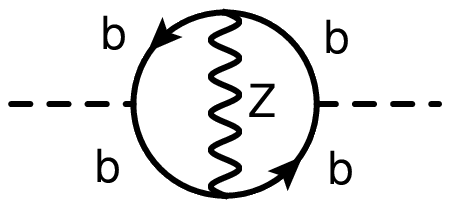} \quad
    \includegraphics[width=.22\textwidth]{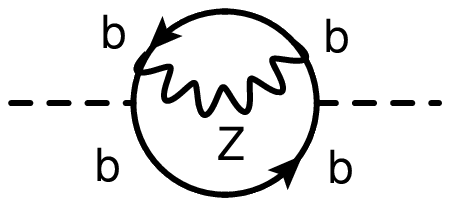} \quad
    \includegraphics[width=.22\textwidth]{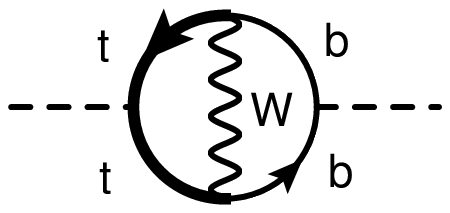} \quad
    \includegraphics[width=.22\textwidth]{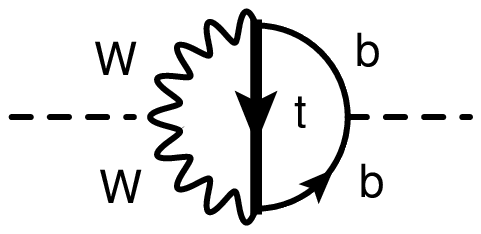} 
    \caption{\label{fig::2loop}Sample Feynman diagram contributing to
      the ${\cal O}(\alpha)$ corrections of $\Gamma(H\to b\bar{b})$.
      External dashed lines denote the Higgs boson.}
  \end{center}
\end{figure}

For the evaluation of the decay rate $\Gamma(H\to b\bar{b})$ we use
the optical theorem which for our application has the form
\begin{eqnarray}
  \Gamma(H\to b\bar{b}) &=& \frac{1}{M_H} 
  \mbox{Im}\left[\Sigma_H(q^2=M_H^2+i\epsilon)\right]
  \,,
\end{eqnarray}
where $\Sigma_H(q^2)$ is the Higgs boson two-point function which is evaluated
on the Higgs boson mass shell.  As a consequence we have to consider
Feynman diagrams as shown in Fig.~\ref{fig::2loop} to evaluate the ${\cal
  O}(\alpha)$ corrections. In this approach we automatically take into account
the on-shell wave function renormalization which is the reason why we have not
considered it in Eq.~(\ref{eq::DeltaalCT}).  Note that we neglect $m_b$
corrections except the leading $m_b^2$ factor and thus the contributing
diagrams either contain $Z$ bosons (possibly together with neutral
Goldstone bosons) or $W$ and/or charged Goldestone bosons and
top quarks. 

We express our final result in terms of the $\overline{\rm MS}$ bottom
quark mass which, as is well known, leads to a better perturbative behaviour
of the decay rate. In this context we briefly want to discuss the tadpole
contributions to the bottom quark propagator (see also discussions in
Refs.~\cite{Hempfling:1994ar,Kniehl:2004hfa}). In fact, besides the diagrams
in Fig.~\ref{fig::2loop} there are also contributions where a closed loop is
connected via a $Z$ or Higgs boson to the bottom quark line, so-called
tadpoles. These contributions are exactly canceled by the on-shell counterterm
contributions to the bottom quark mass. For this reason we drop the tadpoles
in both parts from the very beginning.  Note, however, that after dropping the
tadpole contribution in the counterterm $\delta_{m_b}$, it becomes dependent
on the electroweak gauge parameters $\xi_{W/Z}$. The same is true for the
contribution from the diagrams in Fig.~\ref{fig::2loop}. In the sum $\xi_{W/Z}$
drops out.  In case the bottom quark mass is renormalized in the
$\overline{\rm MS}$ scheme there is no cancellation and the final expression
for $\Gamma(H\to b\bar{b})$ remains $\xi_{W/Z}$-dependent. Note, however, that
also the numerical value of $m_b$ in the $\overline{\rm MS}$ scheme
(formally) depends on
$\xi_{W/Z}$ since in the extraction of $m_b$ from the comparison of theoretical
calculations and experimental data (see, e.g., Ref.~\cite{Chetyrkin:2009fv})
no electroweak tadpoles are included.  The $\xi_{W/Z}$-dependence in
$\Gamma(H\to b\bar{b})$ and $m_b$ cancels.

In our calculation we adopt Feynman gauge in the electroweak sector
but allow for general gauge parameter $\xi_S$ in gluon propagator. In our
final result $\xi_S$ drops out which is a welcome check.
Our Feynman integrals involve the mass scales $M_H, M_t, M_W$ and
$M_Z$. 

Before presenting numerical results let us fix our input parameters
which are given by~\cite{Chetyrkin:2009fv,Agashe:2014kda,Aad:2015zhl}
\begin{eqnarray}
  M_t &=& 173.34~\mbox{GeV}\,,\nonumber\\
  M_H &=& 125.09~\mbox{GeV}\,,\nonumber\\
  M_W &=& 80.385~\mbox{GeV}\,,\nonumber\\
  M_Z &=& 91.1876~\mbox{GeV}\,,\nonumber\\
  m_b(m_b) &=& 4.163~\mbox{GeV}\,,\nonumber\\
  G_F &=& 1.1663788(6)\times10^{-5}~\mbox{GeV}^{-2}\,,\nonumber\\
  \alpha_s(M_Z) &=& 0.1185
\end{eqnarray}
where the four-loop QCD
conversion~\cite{Marquard:2015qpa} of the
on-shell to the $\overline{\rm MS}$ top quark mass leads to
$m_t(m_t)=163.47$~GeV and $m_t(M_H)=166.97$~GeV.

Let us in a first step discuss the contribution from the Feynman diagrams
which do not involve top quarks. As massive particles they only
contain $Z$ bosons  or neutral Goldstone bosons
and thus they depend on $q^2/M_Z^2$ where $q^2=M_H^2$ is
the square of the external momentum. At ${\cal O}(\alpha)$ an exact
calculation is possible, however, at ${\cal O}(\alpha\alpha_s)$ the occurring
integrals become complicated. Thus we evaluate this class of Feynman diagrams
in the limit $q^2\ll M_Z^2$ and apply a Pad\'e approximation to construct an
approximation for the physical limit $q^2=M_H^2$.  In principle one could also
imagine to consider $q^2\gg M_Z^2$.  However, this limit contains decays of
the form $H\to ZZ$ which are kinematically forbidden.  On the other hand, for
$q^2\ll M_Z^2$ we neglect contributions from $H\to Z b\bar{b}$, which are,
however, strongly phase-space suppressed. Furthermore, it is possible to
experimentally distinguish this final state from $H\to b\bar{b}$.
Note that the decay $H\to Z b\bar{b}$ is not
included in the result of Ref.~\cite{Dabelstein:1991ky}.

In the limit $q^2\ll M_Z^2$ we obtain for $\Delta^{(\rm weak,Z)}$
the expansion
\begin{eqnarray}
  \Delta^{(\rm weak,Z)} &=& \sum_{j\ge0} d_{2j}
  \left(\frac{M_H^2}{M_Z^2}\right)^j
  \,\,=\,\, \sum_{j\ge0} D_{2j}
  \,.
  \label{eq::Dj}
\end{eqnarray}
where the coefficients $D_k$ are given in Table~\ref{tab::Dj}.
$\Delta^{\rm (weak,Z)}$ includes all relevant contributions from
$\Delta^{\rm (weak)}_{\rm CT}$ and is thus finite in the limit
$\epsilon\to0$. The bottom quark is renormalized in
the $\overline{\rm MS}$ scheme.
The counterterm contribution is not expanded in
$M_H^2/M_Z^2$ and is contained in the coefficient $D_0$.  Furthermore,
we choose $\mu^2=M_H^2$ for the renormalization scale.

\begin{table}[t]
  \begin{center}
  \begin{tabular}{r|c}
    \hline
    $k$ & $D_k$ \\
    \hline
$0$ & $-0.008768$ \\
$2$ & $-0.000847$ \\
$4$ & $-0.000196$ \\
$6$ & $+0.000269$ \\
$8$ & $-0.000534$ \\
$10$ & $+0.000894$ \\
$12$ & $-0.001515$ \\
$14$ & $+0.002568$ \\
$16$ & $-0.004383$ \\
    \hline
  \end{tabular}
  \caption{\label{tab::Dj}Coefficients $D_k$ from Eq.~(\ref{eq::Dj})
    for $\mu^2=M_H$.}
  \end{center}
\end{table}

\begin{table}[t]
  \begin{center}
  \begin{tabular}{c|c}
    \hline
    Pad\'e approximant & $\Delta^{\rm (weak,Z)}$ \\ \hline
$[3/3]$ & $-0.009744$ \\
$[3/4]$ & $-0.009746$ \\
$[4/3]$ & $-0.009747$ \\
$[4/4]$ & $-0.009746$ \\
\hline
    exact & $-0.009747$\\ \hline
  \end{tabular}
  \caption{\label{tab::Dj_pade}Numerical results for
    $\Delta^{\rm (weak,Z)}$ obtained from the construction
    of Pad\'e approximations using the coefficients in
    Table~\ref{tab::Dj}. The last row contains the exact result
    from Ref.~\cite{Dabelstein:1991ky}.}
  \end{center}
\end{table}

In a next step we use the results in Table~\ref{tab::Dj} and construct various
Pad\'e approximations. The results are shown in Table~\ref{tab::Dj_pade} where
also the exact result for $\Delta^{\rm (weak,Z)}$ from
Ref.~\cite{Dabelstein:1991ky} is displayed.  The deviation of the numerical
approximation based on the $[4/4]$ Pad\'e expression
and the exact result~\cite{Dabelstein:1991ky} is about 0.01\%
which justifies the use of this method at order
$\alpha\alpha_s$.

Let us next turn to the contribution involving top quarks
and a $W$ and/or charged Goldstone bosons.
To simplify the integrals and to obtain simple final expressions
we assume one of the following hierarchies
\begin{eqnarray}
  (A) && M_H^2 \ll 4M_W^2 \ll 4M_t^2\,,\nonumber\\
  (B) && M_H^2 \ll 4M_W^2 \approx 4M_t^2\,.
  \label{eq::hierI}
\end{eqnarray}
We stress that the ${\cal O}(\alpha)$ corrections can be computed without any
assumptions on the relative size of the involved mass scales. However, at
order $\alpha\alpha_s$ the hierarchies in Eq.~(\ref{eq::hierI}) significantly
simplify the calculation.

In Eqs.~(\ref{eq::hierI}) the strong hierarchy (``$\ll$'') means that
we apply an asymptotic expansion~\cite{Smirnov:2002pj} in the
corresponding mass ratio. In the case of an approximation sign we
Taylor-expand the integrand in the mass difference.  As a result we
obtain $\Delta^{\rm (weak)}$ in the form
\begin{eqnarray}
  \Delta^{\rm (weak)} &=& 
  \sum_{i\ge-1} c_{{2i}}^{(A)} \left(\frac{M_W^2}{M_t^2}\right)^i
  \,\,=\,\, 
  \sum_{i\ge-1} C_{{2i}}^{(A)} 
  \,,
  \label{eq::CiIa}
  \nonumber\\
  &=&
  \sum_{i\ge-1} c_{{2i}}^{(B)} \left(\frac{M_t^2-M_W^2}{M_t^2}\right)^i
  \,\,=\,\, 
  \sum_{i\ge-1} C_{{2i}}^{(B)} 
  \,,
  \label{eq::CiIb}
\end{eqnarray}
where the coefficients $c_{k}^{(A)}$ and $c_{k}^{(B)}$ are expansions in
$M_H^2/( 4 M_W^2)$.  Note that by definition $C_0^{(A)}$ and $C_0^{(B)}$
contain the contribution from $\Delta^{\rm (weak,Z)}$.

We show our results for hierarchy $(A)$ in Table~\ref{tab::CiA}
adopting again $\mu^2=M_H^2$ and the $\overline{\rm MS}$ definition for the
bottom quark mass. For the top quark mass both the 
$\overline{\rm MS}$ and on-shell mass value is used.

\begin{table}[t]
  \centering
  \begin{tabular}{r|lc}
    \hline
    $k$ &  $C_{k}^{(A)}$ ($\overline{\rm MS}$)& $\Delta^{\rm (weak)}$ \\ \hline
$-2$ & $+0.005146$ & $+0.005146$ \\
$0$ & $-0.004506  -0.009746|_Z$ & $-0.009106$ \\
$2$ & $-0.000166$ & $-0.009272$ \\
$4$ & $-0.000100$ & $-0.009372$ \\
$6$ & $-0.000105$ & $-0.009477$ \\
$8$ & $-0.000088$ & $-0.009565$ \\
$10$ & $+0.000048$ & $-0.009517$ \\
$12$ & $-0.000029$ & $-0.009546$ \\
$14$ & $+0.000001$ & $-0.009545$ \\
    \hline
    exact && $-0.009549$ \\
    \hline
    \hline
    $k$ &  $C_{k}^{(A)}$ (on-shell)& $\Delta^{\rm (weak)}$ \\ \hline
$-2$ & $+0.004842$ & $+0.004842$ \\
$0$ & $-0.004754  -0.009746|_Z$ & $-0.009659$ \\
$2$ & $-0.000145$ & $-0.009804$ \\
$4$ & $-0.000095$ & $-0.009899$ \\
$6$ & $-0.000078$ & $-0.009977$ \\
$8$ & $-0.000065$ & $-0.010042$ \\
$10$ & $+0.000029$ & $-0.010012$ \\
$12$ & $-0.000018$ & $-0.010031$ \\
$14$ & $+0.000000$ & $-0.010030$ \\
    \hline
    exact && $-0.010034$ \\
    \hline
  \end{tabular}
  \caption{\label{tab::CiA}
    Coefficients $C_{k}^{(A)}$ as defined in Eq.~(\ref{eq::CiIa})
    (middle column). In the $n^{\rm th}$ row of the right column
    the sum including the first $n$ terms is shown.
    On the top part we adopt the $\overline{\rm MS}$ and below
    the on-shell definition for the top quark mass.
    The $Z$ boson contribution is marked by ${}|_Z$.
    }
\end{table}

\begin{figure}[t]
  \begin{center}
    \includegraphics[width=0.9\textwidth]{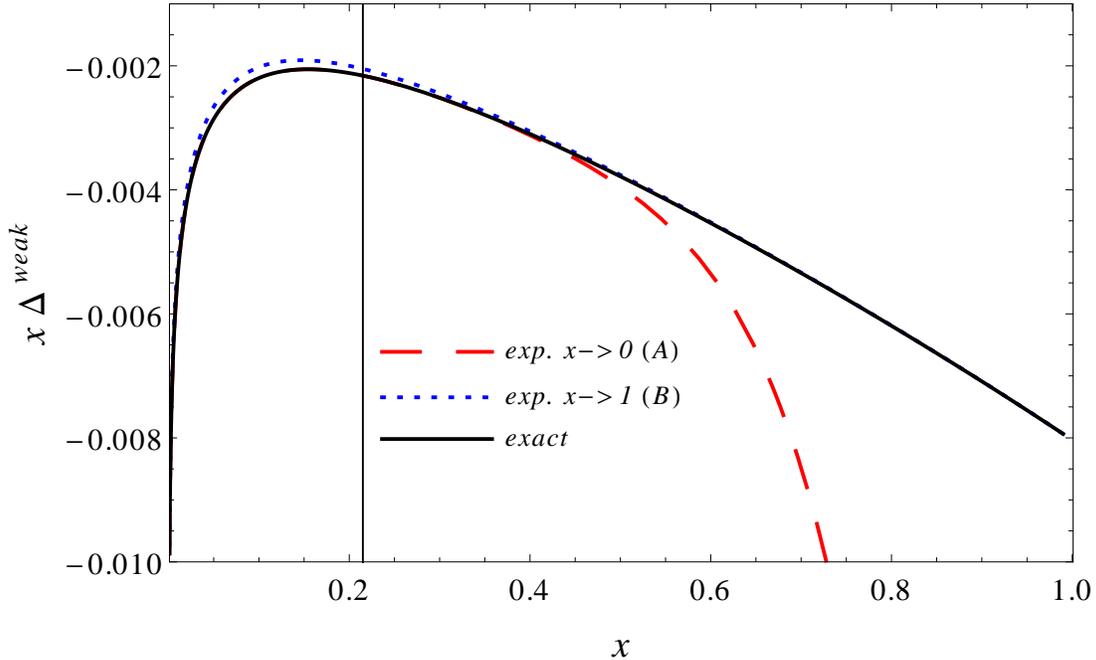} 
    \caption{\label{fig::cmp_with_exact} Comparison of $x \Delta^{\rm (weak)}$
      as obtained for the hierarchies (A) and (B) with the exact result as a
      function of $x=M_W^2/M_t^2$.  The black (solid) curve shows the exact
      result, the (red) dashed curve the expansion for $x\to0$ and the
      (blue) dotted curve the expansion around $x=1$. The vertical line
      indicates the experimental result for $x\approx 0.215$.  For the
      renormalization scale of the bottom quark $\mu^2=M_H^2$ has been chosen.  
      Note that $x \Delta^{\rm (weak)}$ behaves as $\log(x)$ for $x\to0$.} 
    \end{center}
\end{figure}

Note that $\Delta^{\rm (weak)} \sim 1/x$ as $x\to 0$. For this reason we show
in Fig.~\ref{fig::cmp_with_exact} the quantity $x\Delta^{\rm (weak)}$ as a
function of $x=M_W^2/M_t^2$ and compare the expansion obtained for the
hierarchies (A) and (B) with the exact
result~\cite{Dabelstein:1991ky}.\footnote{There is a typo in the quantity
  $\Delta T_{10+11}$ in Eq.(A.2) of~\cite{Dabelstein:1991ky}: the minus sign
  in front of $m_{f^\prime}^2$ should be a plus sign.} For the plot we use the
on-shell definition of the top quark mass and set $\mu^2=M_H^2$.  The numerical
values are obtained by keeping $M_W$ fixed and varying $M_t$.  We take into
account expansion terms up to $k=12$ (see Table~\ref{tab::CiA}) which
corresponds to the expansion depth which is available at order
$\alpha\alpha_s$ (cf. Section~\ref{sec::alas}).  One observes that for $x\lsim
0.4$ a perfect description is obtained from hierarchy (A) (red, dashed curve)
and above $x\approx0.4$ the result from hierarchy (B) (blue, dotted curve)
agrees perfectly with the exact result (black line).  For the physical value
$x\approx 0.215$ one obtains
\begin{eqnarray}
  \Delta^{\rm(weak)}_B &\approx& -0.009525
  \,,
\end{eqnarray}
which has to be compared with the results in Table~\ref{tab::CiA} in the lower
panel. There is a notable  deviation of about 5\% to the exact result 
which has its origin in the divergent behaviour proportional
to $1/x$ for $x\to0$. For this reason we concentrate in
Section~\ref{sec::alas} on hierarchy (A).

\section{\label{sec::alas}Corrections of order $\alpha\alpha_s$}

In this Section we consider the quantity $\Delta^{(\alpha\alpha_s)}$ of
Eq.~(\ref{eq::Gam}). An analytic expression for $\Delta^{(\rm
  QED,\alpha_s)}$ can easily be obtained from the ${\cal O}(\alpha_s^2)$ QCD
corrections (see, e.g., Ref.~\cite{Kataev:1997cq}) after adopting the colour
factors. It reads
\begin{eqnarray}
  \Delta^{(\rm QED, \alpha_s)} \!\!&=& \!
  Q_b^2\frac{\alpha\alpha_s}{\pi^2} C_F \left[
    \frac{691}{32}
    -\frac{3}{4}\pi^2
    -\frac{9}{2}\zeta(3)
    +\frac{105}{8} \ln \ \left(\frac{\mu^2}{M_H^2}\right)
    + \frac{9}{4} \ln ^2\left(\frac{\mu^2}{M_H^2}\right)
  \right]\,.
  \label{eq::emqcd} 
\end{eqnarray}
To obtain $\Delta^{(\rm weak,\alpha_s)}$ we proceed as follows:
\begin{itemize}
\item 
  We consider the imaginary part of the three-loop propagator-type
  diagrams which are obtained by dressing the ${\cal O}(\alpha)$ diagrams 
  (cf. Fig.~\ref{fig::2loop} for examples) in all possible ways with one gluon. 
  This part can be split,
  in analogy to the ${\cal O}(\alpha)$ corrections, into a contribution
  involving $Z$ or Goldstone bosons
  and into a contribution involving $W$ and/or charged Goldstone
  bosons and top quarks.
\item 
  The bare bottom quark mass in the Born result has to be replaced by the
  $\overline{\rm MS}$ renormalized counterpart using corrections of order
  $\alpha\alpha_s$. The corresponding counterterm is available from
  Ref.~\cite{Kniehl:2004hfa,Piclum_DA} 
  which we have checked by an independent calculation.
  It is given by
  \begin{eqnarray}
    \delta_{m_b}^{(\rm weak, \alpha_s), \overline{\rm MS}} &=& C_F
    \frac{\alpha\alpha_s}{\pi^2}\bigg[\frac{1}{\epsilon^2}\left(
      -\frac{1}{16} - \frac{7}{128 c_W^2} -\frac{9}{128 s_W^2}  + \frac{9 m_t^2}{64
        M_W^2 s_W^2}\right)
    \nonumber\\
    &&+  \frac{1}{\epsilon}\left(
      \frac{1}{96} + \frac{31}{768 c_W^2} + \frac{27}{256 s_W^2}  - \frac{3 m_t^2}{32
        M_W^2 s_W^2}
    \right)
    \bigg]\,.
    \label{eq::em2l} 
  \end{eqnarray}
  Note that $\delta_{m_b}^{(\rm weak, \alpha_s), \overline{\rm MS}}$
  contains poles up to order $1/\epsilon^2$ and thus the 
  Born result is needed up to order $\epsilon^2$ terms.
\item 
  $\Gamma^{(0)}\Delta^{(\alpha)}$ has to be available up to order $\epsilon$
  and the bottom and top quark masses have to be
  renormalized using one-loop counterterms of ${\cal O}(\alpha_s)$ which are
  given by
  \begin{eqnarray}
    \delta_{m_q}^{(\alpha_s), {\rm OS}}
    &=& \frac{m_q^{\rm bare}}{M_q} -1 \,\,=\,\, 
      -\frac{\alpha_s}{\pi} 
      C_F  
    \left(\frac{3}{4\epsilon} + 1 +
      \frac{3}{4}\ln\frac{\mu^2}{M_q^2}\right)
    \,,
  \end{eqnarray}
  with $q=b,t$. The corresponding $\overline{\rm MS}$ counterterm
  is obtained by dropping the finite part on the right-hand side
  of the above equation.
\item 
  $\Gamma^{(0)}\Delta^{(\alpha_s)}$ has to be available up to order $\epsilon$
  and the bottom quark mass has to be
  renormalized using one-loop counterterms of ${\cal O}(\alpha)$ which is
  given in Eq.~(\ref{eq::deltaMBalpha})\,.
\item 
  There is a contribution where $v_r$ from Eq.~(\ref{eq::deltaVRa}) multiplies
  $\Gamma^{(0)}\Delta^{(\alpha_s)}$. Since the latter is finite we do not need
  the ${\cal O}(\epsilon)$ part of $v_r$. On the other hand, since $v_r$
  contains $1/\epsilon$ poles $\Gamma^{(0)}\Delta^{(\alpha_s)}$ is needed
  including ${\cal O}(\epsilon)$ terms.
\item 
  The fermion-loop contributions to $v_r$ [see Eq.~(\ref{eq::deltaVRa})]
  receive two-loop QCD corrections which are multiplied by the Born
  decay rate.
  Since the fermionic contribution to $\Sigma^{\gamma Z}(0)$  
  vanishes only $\Sigma^W(0)$ and $\Sigma^{\prime H}(M_H^2)$
  get correction terms of order $\alpha\alpha_s$. We compute them in the
  limit of a heavy top quark and obtain in the $\overline{\rm MS}$ scheme
  \begin{eqnarray}
    \lefteqn{-\frac{\Sigma^W(0)}{M_W^2} - \Sigma^{\prime H}(M_H^2)
      \Bigg|_{{\cal O}(\alpha\alpha_s)} } \nonumber\\
    &=&
    \frac{N_c \alpha \alpha_s}{\pi^2s_W^2}
    \left\{\frac{m_t^2}{M_W^2}\left[\frac{19}{72}
        + \frac{7}{24}\ln\left(\frac{\mu^2}{m_t^2}\right) -
        \frac{1}{12}\zeta(2)\right] -
      \frac{M_H^2}{M_W^2}\frac{61}{3240}\right. 
    \nonumber\\&&
    \left.+ \frac{M_H^4}{m_t^2M_W^2}\left[-\frac{503}{100800} +
        \frac{3}{560}\ln\left(\frac{\mu^2}{m_t^2}\right)\right]
    \right.
    \nonumber\\&&
    \left. 
      + \frac{M_H^6}{m_t^4M_W^2}\left[-\frac{9523}{15876000} +
        \frac{1}{630}\ln\left(\frac{\mu^2}{m_t^2}\right)\right]\right. 
    \nonumber\\&&
    \left.+\frac{M_H^8}{m_t^6M_W^2}\left[-\frac{100687}{2514758400} +
        \frac{1}{2464}\ln\left(\frac{\mu^2}{m_t^2}\right)\right]\right.  
    \nonumber\\&&
    \left. + \frac{M_H^{10}}{m_t^8M_W^2}\left[\frac{154559}{19423404000} +
        \frac{1}{10010}\ln\left(\frac{\mu^2}{m_t^2}\right)\right] +
      \mathcal{O}\left(\frac{M_H^{12}}{m_t^{10} M_W^2}\right)\right\}
    \,,
  \end{eqnarray}
  where $m_t=m_t(\mu)$.
\end{itemize}
The individual terms develop poles up to order $1/\epsilon^2$, which cancel in
the sum.

We are now in the position to present results for the order $\alpha\alpha_s$
corrections. In analogy to Eqs.~(\ref{eq::Dj}) and~(\ref{eq::CiIa})
we introduce
\begin{eqnarray}
  \Delta^{(\rm weak,\alpha_s,Z)} &=& \sum_{j\ge0} d_{2j} 
  \left(\frac{M_H^2}{M_Z^2}\right)^j
  \,\,=\,\, \sum_{j\ge0} D_{2j}
  \,,\nonumber\\
  \Delta^{\rm (weak,\alpha_s)} &=& 
  \sum_{i\ge-1} c_{{2i}}^{(A)} \left(\frac{M_W^2}{M_t^2}\right)^i
  \,\,=\,\, 
  \sum_{i\ge-1} C_{{2i}}^{(A)} 
  \label{eq::Dj_2}
\end{eqnarray}
where for convenience $\Delta^{(\rm weak,\alpha_s,Z)}$ is added
to the coefficient $C_{0}^{(A)}$.

\begin{table}[t]
  \begin{center}
  \begin{tabular}{r|c}
    \hline
    $k$ & $D_k$ \\
    \hline
$0$ & $-0.001692$ \\
$2$ & $-0.000032$ \\
$4$ & $-0.000547$ \\
$6$ & $+0.000836$ \\
$8$ & $-0.001347$ \\
$10$ & $+0.002183$ \\
$12$ & $-0.003604$ \\
$14$ & $+0.006033$ \\
$16$ & $-0.010223$ \\
   \hline
\end{tabular}
  \caption{\label{tab::Dj_2}Coefficients $D_k$ from Eq.~(\ref{eq::Dj_2})
    for $\mu^2=M_H^2$.}
  \end{center}
\end{table}

\begin{table}[t]
  \begin{center}
  \begin{tabular}{c|l}
    \hline
    Pad\'e approximant & $\Delta^{\rm (weak,\alpha_s,Z)}$ \\ \hline
$[3/3]$ & $-0.001955$ \\
$[3/4]$ & $-0.001954$ \\
$[4/3]$ & $-0.001960$ \\
$[4/4]$ & $-0.001953$ \\
\hline
  \end{tabular}
  \caption{\label{tab::Dj_pade_2}Numerical results for
    $\Delta^{\rm (weak,\alpha_s,Z)}$ obtained from the construction
    of Pad\'e approximations using the coefficients in
    Table~\ref{tab::Dj_2}.}
  \end{center}
\end{table}

In the case of $\Delta^{(\rm weak,\alpha_s,Z)}$ we proceed as at order
$\alpha$: we compute nine expansion terms for the (formal) limit
$q^2\ll M_Z^2$ and set $q^2=M_H^2$. After including the corresponding
counterterm contributions we obtain the expansion coefficients listed in
Table~\ref{tab::Dj_2}.  Afterwards we construct several Pad\'e approximants
and obtain the results in Table~\ref{tab::Dj_pade_2}.  We observe a similar
stability as at ${\cal O}(\alpha)$ and estimate the final result as
\begin{eqnarray}
  \Delta^{(\rm weak,\alpha_s,Z)} &=& -0.00195(1)
  \,,
\end{eqnarray}
which has an uncertainty of about $0.5\%$, an accuracy sufficient for all
foreseeable applications.

\begin{table}[t]
  \centering
  \begin{tabular}{r|lc}
    \hline
    $k$ &  $C_{k}^{(A)}$ ($\overline{\rm MS}$)& $\Delta^{\rm (weak,\alpha_s)}$
    \\ \hline
$-2$ & $-0.000479$ & $-0.000479$ \\
$0$ & $-0.000514  -0.001953|_Z$ & $-0.002946$ \\
$2$ & $-0.000044$ & $-0.002990$ \\
$4$ & $+0.000018$ & $-0.002972$ \\
$6$ & $+0.000003$ & $-0.002970$ \\
$8$ & $+0.000005$ & $-0.002964$ \\
$10$ & $+0.000004$ & $-0.002960$ \\
$12$ & $+0.000002$ & $-0.002959$ \\
    \hline
    \hline
    $k$ &  $C_{k}^{(A)}$ (on-shell)& $\Delta^{\rm (weak,\alpha_s)}$ \\ \hline
$-2$ & $-0.000481$ & $-0.000481$ \\
$0$ & $-0.000382  -0.001953|_Z$ & $-0.002816$ \\
$2$ & $-0.000032$ & $-0.002848$ \\
$4$ & $-0.000006$ & $-0.002854$ \\
$6$ & $-0.000008$ & $-0.002862$ \\
$8$ & $+0.000011$ & $-0.002851$ \\
$10$ & $-0.000010$ & $-0.002861$ \\
$12$ & $+0.000002$ & $-0.002860$ \\
    \hline
  \end{tabular}
  \caption{\label{tab::CiA_2}
    Coefficients $C_{k}^{(A)}$ at order $\alpha\alpha_s$ as defined in Eq.~(\ref{eq::Dj_2}).
    In the $n^{\rm th}$ row of the right column                                               
    the sum including the first $n$ terms is shown.
    On the top part we adopt the $\overline{\rm MS}$ and below
    the on-shell definition for the top quark mass.
  }
\end{table}

In Table~\ref{tab::CiA_2} we present the results for the coefficients
$C_{k}^{(A)}$. We observe a continuous decrease of the magnitude 
leading to a stable result with two significant digits after including
six expansion terms, a similar behaviour as at order $\alpha$. 
The seventh and eighth terms confirm this approximation.
It is also interesting to note that the contribution from the $Z$ boson
diagrams amounts to about 65\% of the total result. Furthermore, the leading
$m_t^2$ contribution amounts to less than 20\% of $\Delta^{\rm
  (weak,\alpha_s)}$ but to more than 50\% of the $W$ boson diagrams,
i.e., $\Delta^{\rm (weak,\alpha_s)} - \Delta^{\rm (weak,\alpha_s,Z)}$.


\section{\label{sec::num}Numerical results and conclusions}

\begin{table}[t]
  \centering
  \renewcommand{\arraystretch}{1.2}
  \begin{tabular}{l|cccc}
    \hline
    & $\Delta^{\rm (\alpha_s)}$ & $\Delta^{\rm (\alpha_s^2)}$ &
    $\Delta^{\rm (\alpha_s^3)}$ & $\Delta^{\rm (\alpha_s^4)}$ 
    \\
    QCD & 0.2040 & 0.0378 & 0.0020 & $-0.0014$  \\
    \hline
    &  $\Delta^{\rm (QED)}$& $\Delta^{\rm (QED,\alpha_s)}$& \\
    QED/QCD & 0.0011 & 0.0001 \\
    \hline
    & $\Delta^{\rm (weak)}$& $\Delta^{\rm (weak,\alpha_s)}$ &
    $\Delta^{\rm (weak,Z)}$& $\Delta^{\rm (weak,\alpha_s,Z)}$ \\
    weak/QCD & $-0.0100$ & $ -0.0029$ & $-0.0097$ & $ -0.0020$ \\
    \hline 
  \end{tabular}
  \caption{\label{tab::delta_mu}Numerical result for the QCD, QED and
    weak one-loop and mixed two-loop corrections
    for $\mu^2=M_H^2$. Note that $\Delta^{\rm (weak)}$ and $\Delta^{\rm
      (weak,\alpha_s)}$ contain the contributions from
    $\Delta^{\rm (weak,Z)}$ and $\Delta^{\rm (weak,\alpha_s,Z)}$, respectively.}
\end{table}

In Table~\ref{tab::delta_mu} we summarize our results for the ${\cal
  O}(\alpha)$ and ${\cal O}(\alpha\alpha_s)$ corrections where the electroweak
part is split into QED and weak corrections.  The contribution from the $Z$
boson diagrams is listed for completeness; their contribution is contained in
$\Delta^{\rm (weak)}$ and $\Delta^{\rm (weak,\alpha_s)}$, respectively.  For
comparison also the QCD corrections up to ${\cal
  O}(\alpha_s^4)$~\cite{Baikov:2005rw} based on computations of the
imaginary part of the massless Higgs correlators are shown in
Table~\ref{tab::delta_mu}. Top quark induced QCD corrections due to an
effective $Hb\bar{b}$ coupling, which are in general small (see,
e.g. Eq.~(14) of Ref.~\cite{Liu:2015fxa}), are not shown.

Both at one- and two-loop order the weak corrections are
negative whereas the QED
corrections are positive. One furthermore observes that the weak
corrections are about an order of magnitude larger than the QED terms.  For
$\mu^2=M_H^2$ the weak corrections amount to about $-1\%$ which is significantly
smaller than the one-loop QCD correction ($+20\%$), however, it is of the same
order of magnitude as the two-loop ${\cal O}(\alpha_s^2)$ corrections
obtained
from the massless Higgs correlator, which amount to $+3.8\%$ (see, e.g.,
Ref.~\cite{Baikov:2005rw}).  At the same value of $\mu$ the correction term $\Delta^{\rm
  (weak,\alpha_s)}$ amounts to about $-0.3\%$ which is a factor three larger than the
one-loop QED corrections and which is of the same order of magnitude,
but with the
opposite sign, as the three-loop QCD corrections. It is interesting to
note that the four-loop QCD corrections are $-0.1\%$.

Finally, it is interesting to comment on the assumption the QED and QCD
corrections factorize, an approach often chosen in case ${\cal
  O}(\alpha\alpha_s)$ terms are missing. To do this we define
\begin{eqnarray}
  \Delta^{(\alpha\alpha_s, \rm non-fact.)} &=& \Delta^{(\alpha\alpha_s)}
  - \Delta^{(\alpha)} \Delta^{(\alpha_s)}
  \,,
\end{eqnarray}
which shall be small in case the factorization approach works.
From the numbers in Table~\ref{tab::delta_mu} we obtain
\begin{eqnarray}
  \Delta^{(\alpha\alpha_s, \rm non-fact.)} &=& -0.000831
  \,,
\end{eqnarray}
which corresponds to about 30\% of $\Delta^{(\alpha\alpha_s)}$.

To summarize, in this letter we have computed the complete ${\cal
  O}(\alpha\alpha_s)$ mixed corections to the decay rate $\Gamma(H\to
b\bar{b})$. They provide a negative shift of about $-0.3\%$ to $\Gamma(H\to
b\bar{b})$ which corresponds to about $30\%$ of the one-loop electroweak
corrections and which is of the same order of magnitude as the three-loop
QCD corrections.



\section*{Acknowledgements}

We would like to thank J.H. K\"uhn for many discussions and carefully
reading the manuscript.
This work was supported by the BMBF through 05H15VKCCA.
L.M. is supported by the DFG through Heisenberg Stipendium MI 1358/1-1.


\end{document}